\documentclass[prb,twocolumn]{revtex4}  

\usepackage{graphicx}
\usepackage{color}

\begin{document}
\title{Fictitious excitations in the classical Heisenberg antiferromagnet on the kagome lattice}

\author{Stefan Schnabel}
\thanks{Present address:  Institut f\"ur Theoretische Physik, Universit\"at Leipzig, Germany}
\email{schnabel@itp.uni-leipzig.de}
\affiliation{Center for Simulational Physics, University of Georgia, Athens, GA 30602, USA}
\author{David P. Landau}
\affiliation{Center for Simulational Physics, University of Georgia, Athens, GA 30602, USA}
\date{\today}

\begin{abstract}
Using an advanced Monte Carlo algorithm and high precision spin dynamics simulation, we investigated the dynamical behavior of the Heisenberg antiferromagnet on a kagome lattice at extremely low temperatures. We demonstrate that the detection and correct identification of propagating excitations depends crucially on the choice of coordinates and we show how modes are displaced in Fourier space if a single, global Cartesian coordinate system is chosen. 
\end{abstract}

\maketitle

\section{Introduction}
Although frustrated magnetic systems have enjoyed the attention of the scientific community for many years, they remain fascinating research subjects and their rich behavior is far from being completely understood. If only classical interactions are considered, it is usually an easy task to find the ground state, which is often degenerate. But once the systems are thermally excited, intriguing effects such as vortices\cite{vortices} or magnetic monopoles\cite{monopoles} can be observed.

Computational physics provides important tools for probing the behavior of these magnets. Furthermore, it bridges the gap between theory and experiment by testing theoretical predictions which are usually stated for idealized systems that do not exist in their pure form in nature and are, therefore, not accessible by experiments. On the other hand, more realistic models are often too complex for rigorous analysis but are to some extent treatable in computer simulations, which are becoming increasingly powerful due to improved algorithms and faster machines. Spin dynamics methods are valuable in this respect because they allow the determination of the dynamic structure factor which is closely related to results of inelastic neutron scattering experiments.

One system of great interest is the kagome Heisenberg antiferromagnet (KHAFM). Studies of quantum antiferromagnets helped pique interest in 
corresponding classical systems, see Ref. \onlinecite{kagome_elser,kagome_berlinsky}, and both experiments and numerical studies of quantum systems 
continue to this day \cite{hyperkagome_exp,spin_liquid}. Twenty years ago an extensive theoretical description\cite{kagome_elser,kagome_berlinsky,kagome_shender} was developed  for the classical Heisenberg model, and a number of numerical studies\cite{kagome_huse,kagome_berlinsky2,kagome_mc_zhito} have been performed since. Predictions for dynamical behavior were, in part, confirmed by recently published results of spin dynamics simulations\cite{kagome_sd_prl}. However, a branch of unpredicted modes was found and its origin has not yet been clarified. This makes a closer look at the subject worthwhile.

The paper is organized as follows: After this introduction we will introduce the model and review some established facts about its behavior. In section III, the methods used in our work for simulation and analysis are presented. Section IV is devoted to a presentation and discussion of the results of the Monte Carlo and spin dynamic simulations, followed
by concluding remarks in Section V. The paper concludes with two appendices in which we discuss some calculations in greater detail and introduce a modification of the Wang-Landau algorithm\cite{wanglandau, wanglandau2, wanglandau3}.

\begin{figure}
\begin{center}
\includegraphics[width=.8\columnwidth]{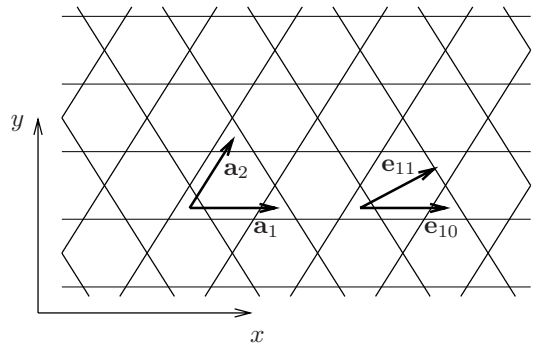}
\caption{\small{\label{fig:lattice} \emph{The kagome lattice and the two lattice vectors $\mathbf a_1$ and $\mathbf a_2$. Spins are placed at the line intersections. The [10] and [11] lattice directions are shown by (unit) vectors $\mathbf e_{10}$ and $\mathbf e_{11}$.}}}
\end{center}
\end{figure}

\section{Background}
In the classical Heisenberg model, three dimensional-vectors $\mathbf s$ of unit length interact via the Hamiltonian
\begin{equation}
\mathcal{H}=-J\sum\limits_{\langle ij\rangle}{\bf s}_i\cdot{\bf s}_j,
\end{equation}
with a negative interaction constant $J$ favoring antiferromagnetic spin pairs. The kagome lattice is drawn in Fig.~\ref{fig:lattice}. We apply periodic boundary conditions and assume in the following that the lattice constant $|\mathbf a_i|=1$. If $N$ is the number of sites, the system contains $2N/3$ triangles.  The energy is minimal ($E_{\rm min}=-N|J|$) if the sum of the three spins belonging to each triangle is zero. This leaves the system underdetermined (frustrated) and the ground state is highly degenerate.

At low enough, but non-zero, temperatures a common spin plane is established, at least in finite systems, in a process called `order by disorder'\cite{kagome_berlinsky2}. This `coplanar' state is stabilized by the entropy of local modes which can be constructed on the hexagons of the kagome lattice. Only three distinct spin direction occur separated by angles of $2\pi/3$ which allows the assignment of labels $\sigma_i\in\{1,2,3\}$ to each spin $\mathbf s_i$ according to the direction towards which it is oriented.

\begin{figure}
\begin{center}
\includegraphics[width=.85\columnwidth]{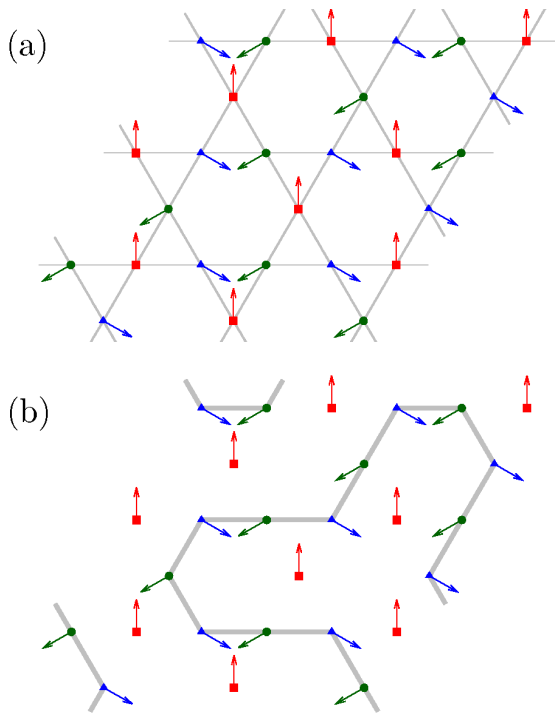}
\caption{\small{\label{fig:cpln_loops} \emph{(a) Typical configuration at low temperatures. Spins fluctuate around three distinct directions which are indicated by different symbols and colors. (b) Solid lines show \{weathervane} loops which are constructed if spins of two 'types' are connected.}}
\end{center}
\end{figure}

\begin{figure}
\begin{center}
\includegraphics[width=.85\columnwidth]{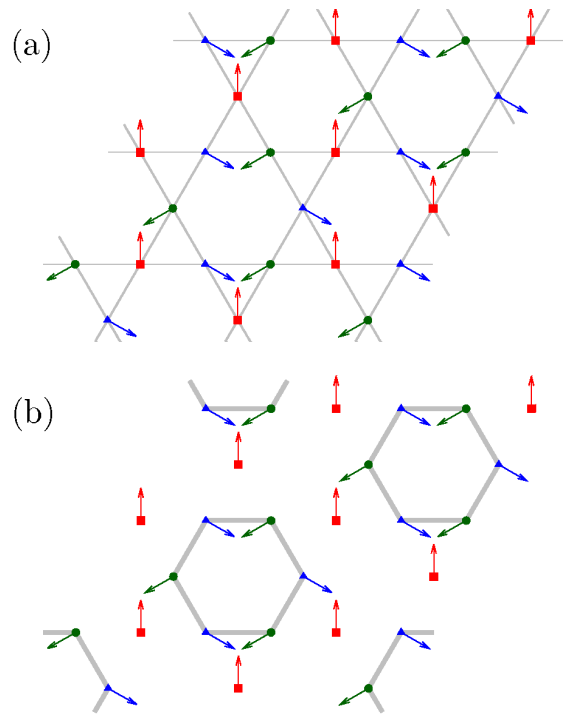}
\caption{\small{\label{fig:cpln_loops33} \emph{(a)The $\sqrt3\!\times\!\sqrt3$ structure; (b) Weather vane loops for the $\sqrt3\!\times\!\sqrt3$ structure are shown by solid lines.}}}
\end{center}
\end{figure}

It is then possible to classify configurations $\{\mathbf s\}$ according to their labels $\{\sigma\}$. The ground state condition requires that each triangle has spins with three different labels; hence, valid $\{\sigma\}$ are also ground state configurations of a three-state Potts model. However, while the probability distribution over all $\{\sigma\}$ is uniform in the case of the Potts model, there is a selection in the KHAFM due to entropic effects. This can be understood if the concept of weathervane loops is employed. These loops form when two label values are chosen and spins possessing these values are connected along the bonds of the lattice (Fig.~\ref{fig:cpln_loops}[b]). Because there are three ways of choosing two out of the three possible values, three overlapping loop structures coexist and each spin belongs to two loops. However, within a single structure, loops never touch but are separated by spins pointing in the third direction. Therefore, spins of a single loop can rotate collectively around this direction while causing only small changes to spin-spin angles and, thus, to energy. However, unlike in the case of a true weather vane complete rotations rarely occur. Instead, relatively large fluctuations in these collective degrees of freedom occur and account for the difference between the ground state of the Potts model and the distribution over different $\{\sigma\}$ in the KHAFM. Because each loop provides one degree of freedom for fluctuations, a structure with many loops has a higher entropy than a structure with few loops. In a counter effect, however,  entropy decreases if the loop number becomes very large, and the maximum loop number does not occur in large enough systems, because it is realized in only six distinct configurations $\{\sigma\}$ possessing the so-called $\sqrt3\!\times\!\sqrt3$ structure (Fig.~\ref{fig:cpln_loops33}). In the $\sqrt3\!\times\!\sqrt3$ state each hexagon is a weathervane loop and the loop structure is periodic in three directions with the six wave vectors $\mathbf k_{\sqrt3\!\times\!\sqrt3}=\pm \frac83\pi\mathbf a_1,\,\pm \frac83\pi\mathbf a_2,\,\pm \frac83\pi(\mathbf a_1-\mathbf a_2)$.

In an analytical study, \cite{kagome_berlinsky} Harris et al. examined the KHAFM with and without- second and third-nearest-neighbor interactions and predicted a branch of soft modes with frequencies approaching zero as the temperature decreases as well as a two-fold degenerate acoustic branch. In the case of only nearest neighbor interactions the frequencies of the acoustic branch were given by
\begin{equation}
\omega(\mathbf k)=|J|\sqrt{2\left(\sin^2q_1+\sin^2q_2+\sin^2(q_1-q_2)\right)},
\label{eqn:freq_general}
\end{equation}
where $q_1=k_x$, $q_2=(k_x-\sqrt3k_y)/2$, and $\mathbf k$ is the wave vector.

\section{Methods}
\subsection{Alternative Coordinates}
\begin{figure}
\begin{center}
\includegraphics[width=.6\columnwidth]{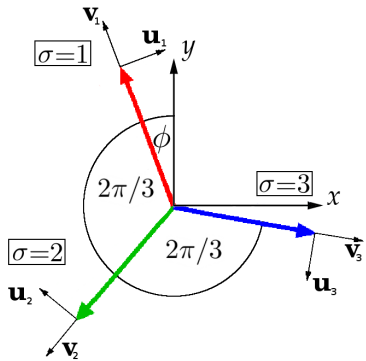}
\caption{\small{\label{fig:in_plane_coordinates} \emph{With decreasing temperature spins tend towards three basic directions and labels, $\sigma$, can be assigned to each spin accordingly. Fluctuations around these directions are described in the alternative coordinates $u,v$ given in Eq. \ref{fig:in_plane_coordinates}.}}}
\end{center}
\end{figure}

Because no anisotropy and no external field are considered, the Hamiltonian has spherical symmetry in spin space and the system can be freely rotated. In the following, we will choose the orientation in such a way that the spin plane coincides with the $xy$-plane,
and we name the angle between the average direction of the $\sigma=1$ spins and the $y$-axis $\phi$. Alternative in-plane coordinates $s^u,s^v,s^z$ as used by Harris et al.\cite{kagome_berlinsky} are now given by
\begin{equation}
      \left(\begin{array}{c}s_i^u\\s_i^v\\s_i^z\end{array}\right)=\left(\begin{array}{ccc}\cos\psi_i & \sin\psi_i & 0\\ -\sin\psi_i & \cos\psi_i & 0\\0 & 0 & 1\end{array}\right) \left(\begin{array}{c}s_i^x\\s_i^y\\s_i^z\end{array}\right)
,
\label{eqn:coord_trans}
\end{equation}
where $\psi_i=\phi+2\pi/3(\sigma_i-1)$ is the angle enclosed by the $y$-axis and the basic spin direction which is  denoted by $\sigma_i$. The new $u$-axis is perpendicular and the $v$-axis is parallel to this direction (Fig.~\ref{fig:in_plane_coordinates}).

\subsection{Monte Carlo simulations}
We used the simulated tempering \cite{simulated_temp} technique which allowed us to cover a large temperature range. It provides a generalized ensemble that is a composition of $N_T$ canonical ensembles with a priori defined temperatures $T_i$. The probability to find the system in a  configuration $\{\mathbf s\}$ at temperature $T_i$ is
\begin{equation}
p_i\left(\{\mathbf s\}\right)=\frac{W_i}{Z_{\rm ST}}\exp\left(-\frac{\mathcal{H(\{\mathbf s\})}}{k_{\rm B}T_i}\right),
\label{eqn:prop_ST}
\end{equation}
where $W_i$ is a weight factor and the normalization constant $Z_{\rm ST}$ is the partition sum of the entire ensemble
\begin{equation}
Z_{\rm ST}=\sum\limits_{i=1}^{N_T} W_i Z(T_i)
\end{equation}
which is here written as a weighted sum of canonical partition sums $Z$. The acceptance probability for a Monte Carlo step from configuration $\{\mathbf s\}$ to configuration $\{\mathbf s'\}$ and temperature $T_i$ to temperature $T_j$ is given by
\begin{equation}
P_{ij}^{\rm acc}(\{\mathbf s\},\{\mathbf s'\})=\min\left(1,\frac{W_j}{W_i}\exp\left(\frac{\mathcal{H(\{\mathbf s\})}}{k_{\rm B}T_i}-\frac{\mathcal{H(\{\mathbf s'\})}}{k_{\rm B}T_j}\right)\right)
\label{eqn:p_acc}
\end{equation}

It follows from Eq. (\ref{eqn:p_acc}) that conformational updates can be performed just like in a Metropolis simulation as long as the temperature remains unchanged. Usually, in order to move in temperature space, separate Monte Carlo steps which do not alter the configuration of the system are employed. To ensure that all temperatures are sampled with equal probability, the weights have to be chosen such that $W_i\propto Z(T_i)^{-1}$. Since the partition function is not known at the beginning, we performed preliminary simulations to determine $W_i$ using a modification of the Wang-Landau algorithm that is described in Appendix A. For our simulation, we used $N_T=10000$ different temperatures distributed equidistantly on a logarithmic temperature scale:  $-6\le \log_{10}(k_{\rm B}T_i/|J|)\le 3$. To update configurations at constant temperature, we mainly used the heat-bath \cite{landau_binder} method; in doing so, we avoided the problem of too low or too high acceptance rates as result of improperly chosen step length. Details of the heat-bath technique for a Heisenberg spin system have been published by Miyatake et al.\cite{hbspin_heatbath}.

Obtaining thermodynamic quantities from simulated tempering simulations is relatively simple. Although the reweighted histogram technique \cite{multihistrew} could be applied, it is not required here. It follows from Eq. (\ref{eqn:prop_ST}) that the distribution $P(E)$ can be written in terms of a density of states $g(E)$:
\begin{equation}
P(E)\propto g(E)\sum\limits_{i=1}^{N_T} W_i\exp\left(-\frac{E}{k_{\rm B}T_i}\right).
\end{equation}
Once the weight factors are chosen it is easy to calculate the sum on the right  hand side, and the density of states can be estimated via a histogram over $E$. 

\begin{figure}
\begin{center}
\includegraphics[width=.85\columnwidth]{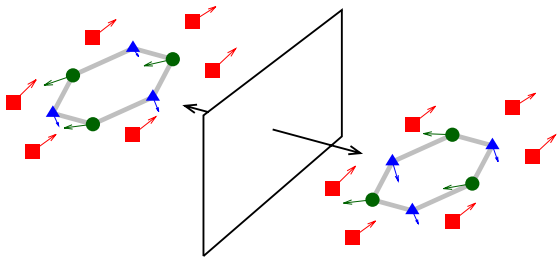}
\caption{\small{\label{fig:loop_upd} \emph{Spins of a weather vane loop can be updated collectively in a reflection. The boundary spins (red) remain unchanged. The mirror plane (rectangle) contains the average direction of the boundary spins and is perpendicular to the (local) spin plane.}}}
\end{center}
\end{figure}

As described above, the coplanar state is highly degenerate. Different configurations $\{\sigma\}$ exist in great number and, although requiring no increase in energy, transitions between them  are inhibited by entropic bottlenecks. Hence, autocorrelation times are large in the coplanar state and ergodicity might even be broken at very low temperatures. This is partially corrected for by use of a generalized ensemble method, such as simulated tempering in our case, but further improvement is possible if jumps between different configurations ${\sigma}$ are enabled at low $T$. To do this, we introduce a Monte Carlo step that is designed to ``flip'' all spins along a weathervane loop. Here, ``flipping'' means exchanging spin directions $\sigma$ that already occur in the loop, i.e., if a weather vane loop is composed of spins with numbers $l_i, i\le N_{\rm loop}$ and $\sigma_{l_i}=1$ or $\sigma_{l_i}=2$ for all spins in the loop, then in the modified loop it is still $\sigma'_i=1$ or $\sigma'_i=2$ for all $i$  but at exchanged positions $\sigma_i\ne\sigma'_i$ (Fig.\ref{fig:loop_upd}). A local reference frame is established with the use of the loop spins and the neighboring (boundary) spins $b_i$, for which in this example $\sigma_{b_i}=3$.

The first step is the identification of a weather vane loop. While this can be done on a global scale by determination of the configuration $\{\sigma\}$, we used a more flexible procedure that, in principle, can result in accepted updates even if the system is in a disordered state. In the beginning, one spin is randomly chosen as the first spin of the loop $\mathbf s_{l_1}$ and one of its neighbors is -- again randomly -- drawn as the second $\mathbf s_{l_2}$. Their common neighbor is recognized as a boundary spin $\mathbf s_{b_1}$ which is not, and cannot be, part of the loop. To continue the construction of the loop, the following procedure is repeated until the loop is closed: Assuming that $n$ spins are part of the incomplete loop, then one neighbor of $\mathbf s_{l_n}$ is part of the loop ($\mathbf s_{l_{n-1}}$) and one neighbor is the boundary spin $\mathbf s_{b_{n-1}}$. The next spin ($\mathbf s_{l_{n+1}}$) must, thus, be chosen from the two remaining neighbors $\mathbf s_{m_1}$ and $\mathbf s_{m_2}$, which are neighbors themselves. Given the nature of a weather vane loop, this spin has to be oriented in approximately the same direction as $\mathbf s_{l_{n-1}}$, so $l_{n+1}={m_1}$ if $\mathbf s_{l_{n-1}}\cdot \mathbf s_{m_1}> \mathbf s_{l_{n-1}}\cdot \mathbf s_{m_2}$ and $l_{n+1}={m_2}$ otherwise. The spin that is not chosen is consequently the boundary spin $\mathbf s_{b_{n}}$. If during this procedure a spin is included that is already a boundary spin or part of the loop but not $\mathbf s_{l_1}$, the procedure has failed and the Monte Carlo move is rejected.

There are two intuitive possibilities to update the loop. The first one is a rotation of the spins of the loop in spin-space by an angle of $\pi$ around the direction given by the boundary spins. This, however, also alters the orientation of the spins with respect to the spin plane. A spin that is pointing above the plane will afterwards point below it and vice versa. This is not desirable because each spin also belongs to a second loop, and the out-of plane component of the spin coordinates is mainly determined by the collective excitation of each loop. Hence, a switch in this component distorts the second loop and increases the energy. One has to choose the alternative, which can be described as a reflection on a plane spanned by the direction of the boundary spins and the normal vector on the spin plane leaving the out-of plane component untouched.

If a closed loop could be constructed, then two vectors $\mathbf v_b,\mathbf v_l$ spanning the local spin plane could easily be found. The first one is the sum of all boundary spins
\begin{equation}
	\mathbf v_b=\sum\limits_{i=1}^{N_{\rm loop}}\mathbf s_{b_i}
\end{equation}
 and a roughly perpendicular second vector is created by alternating the addition and subtraction of the loop spins
\begin{equation}
	\mathbf v_l=\sum\limits_{i=1}^{N_{\rm loop}/2}\mathbf s_{l_{2i}}-\mathbf s_{l_{2i-1}}.
\end{equation}
A normal vector of the plane on which the spins are reflected is then given by 
\begin{equation}
	\mathbf u=\mathbf v_b \times (\mathbf v_b \times \mathbf v_l),
\end{equation}
and the updated spins are
\begin{equation}
	\mathbf s'_{l_i}=\mathbf s_{l_i}-2\mathbf u\frac{\mathbf s_{l_i}\cdot\mathbf u}{|\mathbf u|^2}.
\end{equation}
After each loop flip we must test if the above procedure would also identify the flipped loop. If this is not the case, the inverse move is impossible and the update has to be rejected. The spin plane and the mirror plane will always be identified correctly in the inverse update. Finally, the change in energy has to be determined, and the move is accepted with the probability given in Eq.(\ref{eqn:p_acc}).

\begin{figure}
\begin{center}
\includegraphics[width=\columnwidth]{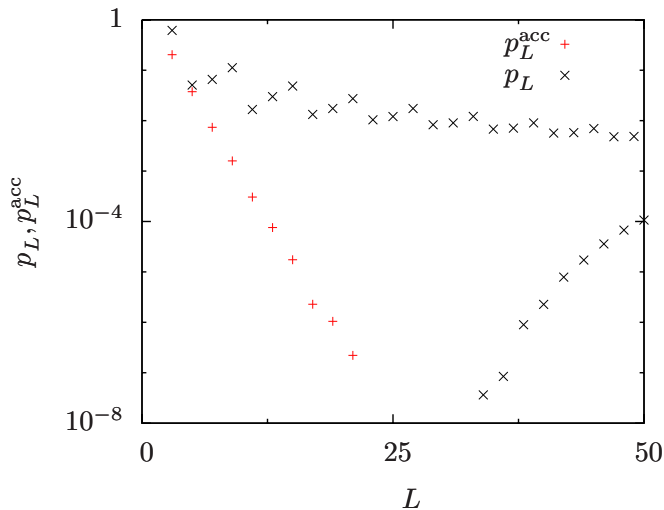}
\caption{\small{\label{fig:loop_distr} \emph{Probability $p_L$ that a constructed loop is of length $L$, and probability $p_L^{\rm acc}$ that an attempted flip of a loop of length $L$ is accepted. Data in the lower right represent loops that span the entire system which for this case ($24^2\times 3$ spins) requires $L\ge24$. (Note that the number of spins in a loop $N_{\rm loop}$ is twice its length $L$ in units of $|\mathbf a_i|$.)}}}
\end{center}
\end{figure}

It turns out that in the coplanar state, the acceptance rates for this update are independent of temperature and decrease exponentially with increasing loop length. This suggests that the average change in energy is positive and proportional to the loop length.

\subsection{Spin dynamics simulations}
Employing the described Monte Carlo algorithm, we extracted configuration $\{\mathbf s\}$ from canonical ensembles at temperatures $T_i$. The coupled equations of motion 
\begin{equation}
\dot{\mathbf s_i}(t)=JH_i(t)\times\mathbf s_i(t)
\end{equation}
were then integrated over a time interval of $t=2000|J|^{-1}$ using a 4th order Adams-Bashforth--predictor--Adams-Moulton--corrector method\cite{Adam-BashMoul} with a time step of $\Delta t=10^{-4}|J|^{-1}$. Here, the local magnetic field $H_i$ is the sum of the spins adjacent to $\mathbf s_i$. The resulting trajectory is subsequently used to calculate space displaced and time displaced correlation functions, e.g., correlations of the spin's $x$-component:
\begin{equation}
      C^x_{\mathbf r}(t)=\left\langle  {s^x}_{\mathbf r'}(0)\cdot{s^x}_{{\mathbf r'}+{\mathbf r}}(t) \right\rangle_{{\mathbf r'}},
\end{equation}
where $\mathbf r'$ runs over all lattice sites. Finally, a space-time Fourier transform results in the dynamic structure factor
\begin{equation}
      S^x({\mathbf k},\omega) = \sum\limits_{{\mathbf r}}\int dt\, C^x_{\mathbf r}(t) e^{-i\mathbf{k}\cdot\mathbf{r}}e^{-i\omega t}
\end{equation}
($S^y,S^z$ similarly), which in turn is closely related to outcomes of inelastic neutron scattering experiments, if global coordinates ($x,y,z$) are used. For the KHAFM, the more genuine description, however, is given by the alternative coordinates ($u,v,z$) and we also calculated the respective structure factors ($S^u,S^v$).

Spin dynamics methods have proven to be valuable for the study of diverse dynamic excitations, e.g., spin-waves\cite{spinwaves}, vortices\cite{vortices}, and solitons\cite{solitons}.

\section{Results and Discussion}
\subsection{Static Properties}
\begin{figure}
\begin{center}
\includegraphics[width=\columnwidth]{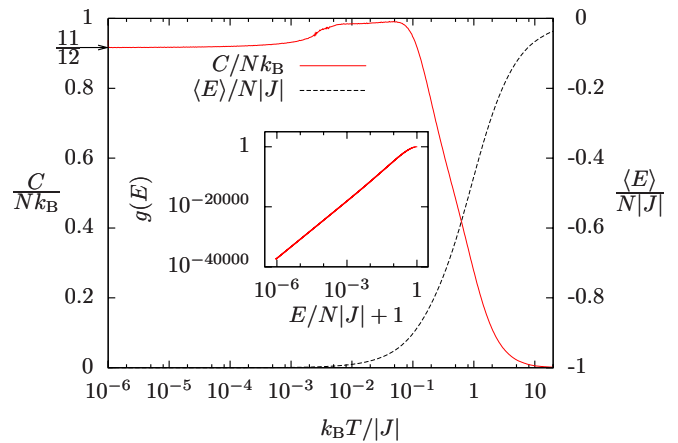}
\caption{\small{\label{fig:sph_dos} \emph{Specific heat $C/N$, mean energy $\langle E\rangle/N$, and density of states $g(E)$ for a system containing $N=6912$ ($=48^2\times3$) spins. Statistical errors do not exceed the line width.}}}
\end{center}
\end{figure}

As a first result we obtained the density of states $g(E)$ and calculated the specific heat (Fig.~\ref{fig:sph_dos}). At low $T$ on each of the $N/3$ hexagons a local soft mode can be constructed. These modes experience anharmonic potentials of fourth order with a thermal energy of $\frac14k_{\rm B}T$. All other degrees of freedom contain $\frac12k_{\rm B}T$, leading to a total average energy of $\langle E\rangle/N|J|=\frac{11}{12}k_{\rm B}T-1$ at small $T$. As in previous studies, \cite{kagome_berlinsky2,kagome_shender,kagome_mc_zhito} this theoretical prediction characterizing the coplanar state is fulfilled for $T\lesssim3\cdot10^{-3}|J|/k_{\rm B}$.

At intermediate temperatures, $C/Nk_{\rm B}\approx1$ and low $E$ indicate that only short range correlations exist and that spherical symmetry is not yet broken in favor of a common spin plane.

Note the linear shape of the density of states in the double-logarithmic plot in the inset of Fig.~\ref{fig:sph_dos}. Although of minor relevance here, this behavior is a general feature of classical systems corresponding to a finite limit of the specific heat for $T\rightarrow0$. This can be shown easily using the following ansatz for the density of states:
\begin{equation}
\log\frac{g(E)}{G}=\gamma\log\left((E-E_0)/|J|\right),
\end{equation}
where $E_0$ is the ground-state energy $E_0=-N|J|$ and $G$ is a normalization constant. One finds that the specific heat $C=\gamma k_{\rm B}$ and the mean energy $\langle E\rangle = E_0+\gamma k_{\rm B}T$. A fit of the data in Fig.~\ref{fig:sph_dos} for ${\log_{10}k_{\rm B}T/|J|<-5.5}$ gives $\gamma=6335.9(2)$ which is close to the value \mbox{$N\cdot11/12=6336$ which we expect for $T\rightarrow0$.}

\begin{figure}
\begin{center}
\includegraphics[width=0.85\columnwidth]{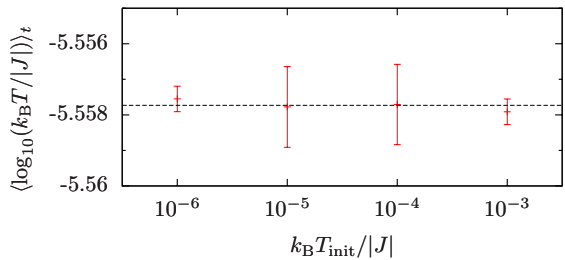}
\caption{\small{\label{fig:av_bta} \emph{If the simulation is constrained to $\log_{10}(k_{\rm B}T/|J|)\le -5$ the time-averaged logarithmic temperatures are independent of the original temperatures $T_{\rm init}$. The broken line is a guide for the eye.}}}
\end{center}
\end{figure}

From simulation and theory, \cite{kagome_berlinsky,kagome_huse,kagome_berlinsky2} it is known that the probability distribution over different configurations $\{\sigma\}$ is non-uniform and biased toward the ${\sqrt3\!\times\!\sqrt3}$ state. It has been under discussion whether this distribution changes with temperature and if long-range order is established for $T\rightarrow0$. To investigate this possibility and to test the algorithm, we produced thousands of conformations of a 432 spin system at different temperatures, cooled them below $T=10^{-5}|J|/k_{\rm B}$, and performed simulated tempering simulations without loop flip updates at $10^{-6}\le k_{\rm B}T/|J|\le 10^{-5}$. Due to the low temperature, in more than half of the simulations no loop flips occurred spontaneously, which enabled a comparison of average logarithmic temperatures of single configurations $\{\sigma\}$. If certain configurations are favored at low temperatures, then this would result in values closer to the lower bound while configurations taken at higher $T$ would consequently show higher average logarithmic temperatures. The results are shown in Fig.~\ref{fig:av_bta}, and within the precision achieved no evidence for such a trend can be found. The exact values have no physical meaning; with perfectly tuned weights $W_i$ the distribution over the logarithmic temperature would be flat and the averages would converge to 5.5.

We conclude that for $\log_{10}(k_{\rm B}T/|J|\le-3$ the ensemble of loop structures is 
practically independent of temperature. Consequently, no further fundamental order is established and changes in  energy, sub-lattice magnetization, or other order parameters \cite{kagome_mc_zhito} are based solely on decreasing excitations on a fixed set of ground-state configurations.

It should be noted, however, that Henley \cite{kagome_henley} recently studied effective Hamiltonians and suggested that long range order may exist for $T\to 0$ at length scales that exceed the size of the systems we could investigate and which could, therefore, be neither confirmed nor excluded by our simulations.

\begin{figure}
\begin{center}
\includegraphics[width=\columnwidth]{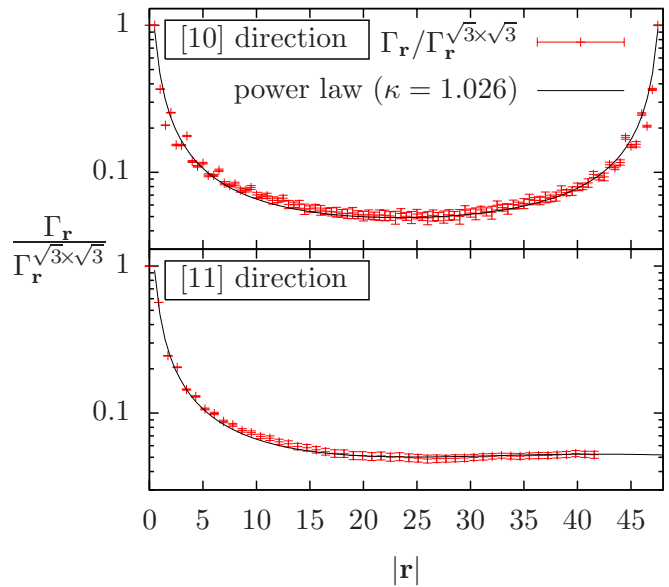}
\caption{\small{\label{fig:sigma_correl} \emph{Static correlations $\Gamma_{\mathbf r}$ for a $48^2\times3$ lattice follow a decaying $\Gamma^{\sqrt3\!\times\!\sqrt3}_{\mathbf r}$ pattern if the system occupies the coplanar state. The solid lines show a power law $\propto|\mathbf r|^{-1.026}$ with included periodic boundary conditions. (In [11] direction $\Gamma_{\mathbf r}$ has  a period of $\sqrt3\cdot48\approx83.1$ )}}}

\end{center}
\end{figure}

It is instructive to investigate the static correlations between labels $\sigma$, which we define as:
\begin{equation}
\Gamma_{{\mathbf r}_i-{\mathbf r}_j}=\langle \delta_{\sigma_i,\sigma_j}\rangle-\frac13.
\end{equation}
This function bears no noticeable temperature dependence as long as the system is in the coplanar state. This confirms once more that the distribution over the loop structures does not change with $T$. Yet, $\Gamma_{\mathbf r}$ is best measured at lower temperatures where $\{\sigma\}$ can be determined most easily. As previous studies observed, we find that the systems show $\sqrt3\!\times\!\sqrt3$ correlations which decrease with increasing distance. The data are well described by a power law:
\begin{equation}
\Gamma_{\mathbf r}=\Gamma^{\sqrt3\!\times\!\sqrt3}_{\mathbf r}|\mathbf r|^{-\kappa},
\label{eq:label_corr}
\end{equation}
where
\begin{equation}
\Gamma^{\sqrt3\!\times\!\sqrt3}_{\mathbf r}=\frac23\cos(\mathbf r\mathbf k_{\sqrt3\!\times\!\sqrt3}).
\label{eq:label_corr33}
\end{equation}
To estimate the exponent $\kappa$, we fitted a power-law function $G$ with included periodic boundary conditions, 
\begin{equation}
G(\mathbf r)=G_0\sum\limits_i\sum\limits_j |\mathbf r - iL\mathbf a_1 - jL\mathbf a_2|^{-\kappa}
\label{eq:pbc_power}
\end{equation}
to $\Gamma_{\mathbf r}/\Gamma^{\sqrt3\!\times\!\sqrt3}$ and obtained $\kappa\approx1$. Some of the relevant data and the fit for a system with $L=48$ are shown in Fig.~\ref{fig:sigma_correl}.

\subsection{Dynamic properties}
In our investigation, we consider solely wave vectors parallel and perpendicular to the bonds of the system, which simplifies Eq.~(\ref{eqn:freq_general}) to
\begin{equation}
\omega(l\mathbf e_{10})=|J|\sqrt{3-2\cos l-\cos 2l},
\label{eqn:freq_10}
\end{equation}
with the unit vector $\mathbf e_{10}=\mathbf a_1,\,\mathbf a_2,\,\mathbf a_1-\mathbf a_2$ and 
\begin{equation}
\omega(l\mathbf e_{11})=2|J|\left|\sin\left( {\sqrt3}l/2\right)\right|,
\label{eqn:freq_11}
\end{equation}
with $\mathbf e_{11}=\frac{\mathbf a_1+\mathbf a_2}{|\mathbf a_1+\mathbf a_2|},\,\frac{2\mathbf a_1-\mathbf a_2}{|2\mathbf a_1-\mathbf a_2|},\,\frac{-\mathbf a_1+2\mathbf a_2}{|-\mathbf a_1+2\mathbf a_2|}$ (Fig.~\ref{fig:lattice}).

All data presented in this section are from simulations of a system containing $48^2\times3$ spins at $T=10^{-6}|J|/k_{\rm B}$. We produced 2457 independent configurations and performed spin dynamics simulations of length $t_{\rm SD}=2000|J|^{-1}$ with a step length of $\Delta t=10^{-4}|J|^{-1}$. Cauchy distributions were fitted to the maxima, and the jackknife method was employed to estimate statistical errors in peak positions. While we will focus on the modes of the acoustic branch, we acknowledge that prominent signals at $\omega=0$ give strong evidence for soft modes.

We also obtained results for smaller lattices and observed the systematic variation in the dynamic structure factor with lattice size. The results for smaller lattices were useful for the eventual interpretation of the large lattice data but do not themselves show different phenomena. For this reason, we do not show those data here.

\begin{figure}
\begin{center}
\includegraphics[width=\columnwidth]{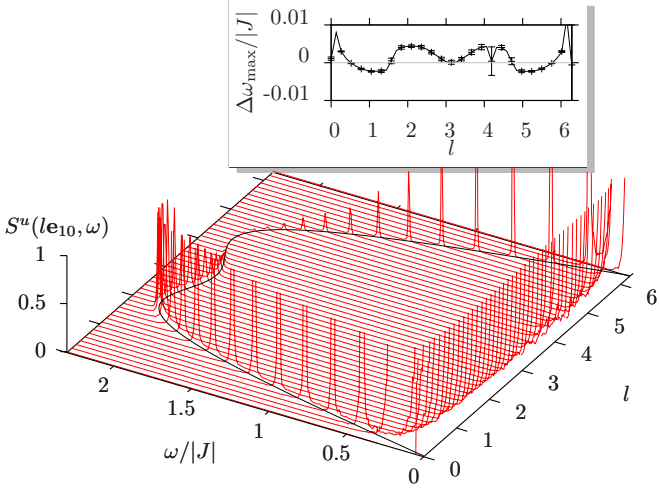}
\caption{\small{\label{fig:10_rel} \emph{Dynamic structure factor in the (10) direction based on the alternative in-plane coordinate $u$. Maxima agree well with the dispersion relation from Eq. (\ref{eqn:freq_10}) which is plotted as a line in $l\omega$-plane. Due to the low temperature systematic deviations (inset) are very small. For the sake of convenience, we increased $S$ by a factor of $10^3$ here and in Figs.~\ref{fig:11_rel},\ref{fig:10_zero}, and \ref{fig:11_zero}.}}}
\end{center}
\end{figure}

\begin{figure}
\begin{center}
\includegraphics[width=\columnwidth]{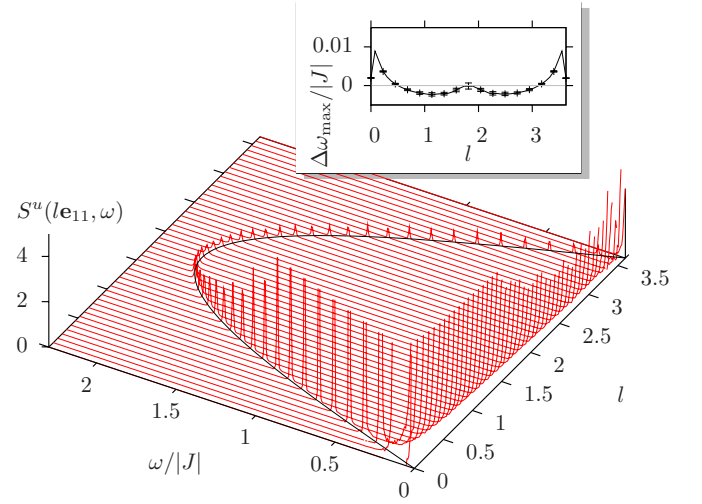}
\caption{\small{\label{fig:11_rel} \emph{Dynamic structure factor for the alternative in-plane coordinate $u$ in the (11) direction. The behavior is well described by the dispersion relation in Eq.~(\ref{eqn:freq_11}) (solid curve). Deviations of peak positions are small (inset).)}}}
\end{center}
\end{figure}

We find that the results for alternative coordinates $u,v,z$ agree very well with the predictions. In both directions, the soft modes produce prominent peaks at very low frequencies. While no trace of acoustic modes can be found in $S^v$, prominent maxima exist in $S^u$ (Figs.~\ref{fig:10_rel},\ref{fig:11_rel}) and $S^z$ at the frequencies given by Eqns.~(\ref{eqn:freq_10},\ref{eqn:freq_11}).

\begin{figure}
\begin{center}
\includegraphics[width=\columnwidth]{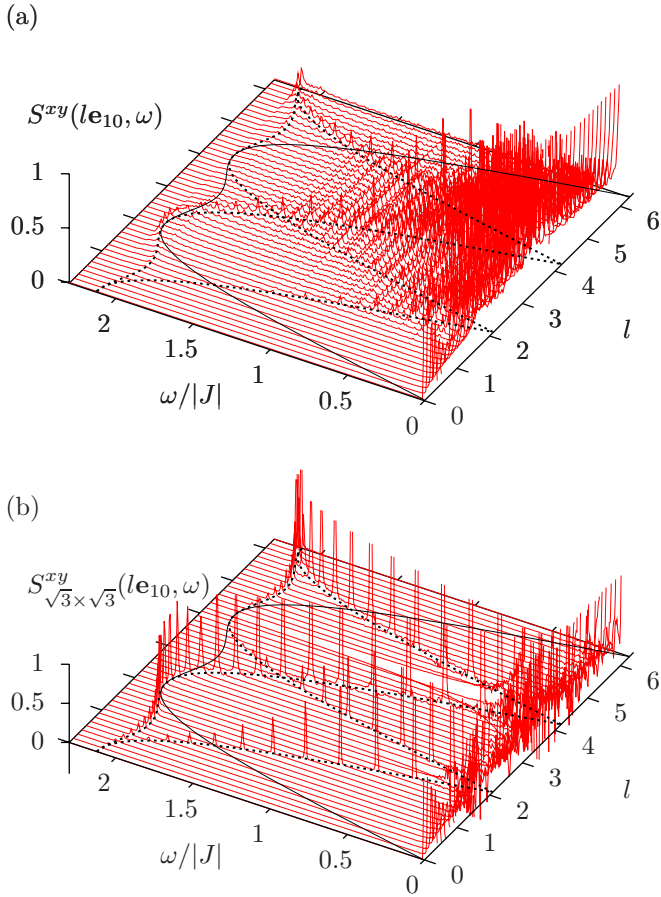}
\caption{\small{\label{fig:10_zero} \emph{Dynamic structure factor in the (10) direction based on global in-plane coordinates.  A shift in Fourier space occurs (see text) and no signals matching the analytical dispersion relation (continuous lines) are observed. In the canonical ensemble (a) a washboard-like pattern is caused by the power law-like decay of the $\sqrt3\!\times\!\sqrt3$ correlations; (b) In case of long-range correlations, e.g. for the pure $\sqrt3\!\times\!\sqrt3$ structure, this pattern vanishes.}}}
\end{center}
\end{figure}

The in-plane dynamic structure factor in global coordinates $S^{xy}=S^x+S^y$, however, shows a completely different behavior. In the (10) direction (Fig.~\ref{fig:10_zero}[a]) arcs of maxima intersect, but no excitations of comparable intensity occur at the frequencies of the acoustic branch according to Eq. (\ref{eqn:freq_10}). In the (11) direction an almost dispersionless branch of maxima with high frequencies resembles an optical mode and was, in fact, interpreted as such by Robert et al. \cite{kagome_sd_prl}.

Additionally, numerous ridges of low intensity form washboard-like patterns in $S^{xy}$ in both directions, but these vanish if a pure $\sqrt3\!\times\!\sqrt3$ structure is considered (Fig.~\ref{fig:10_zero}[b]).  This washboard pattern is pronounced for small lattices but washes out as the lattice size increases.

\begin{figure}
\begin{center}
\includegraphics[width=\columnwidth]{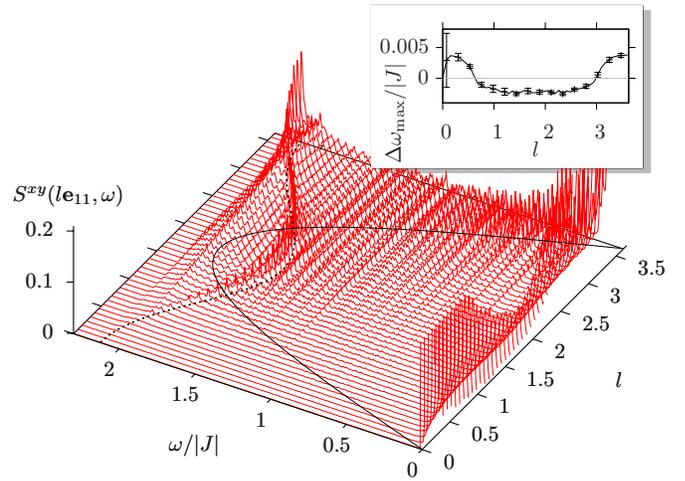}
\caption{\small{\label{fig:11_zero} \emph{As in Fig.~\ref{fig:10_zero}, a shift in Fourier space can also be observed in the (11) direction, and modes from a different part of the Brillouin zone manifest at frequencies given by Eq. (\ref{eqn:freq_11_shift}), drawn as a broken line.
Peak positions used for calculating the deviations in the inset were determined from simulations of the $\sqrt3\!\times\!\sqrt3$ structure allowing a higher precision.}}}
\end{center}
\end{figure}

\begin{figure}
\begin{center}
\includegraphics[width=\columnwidth]{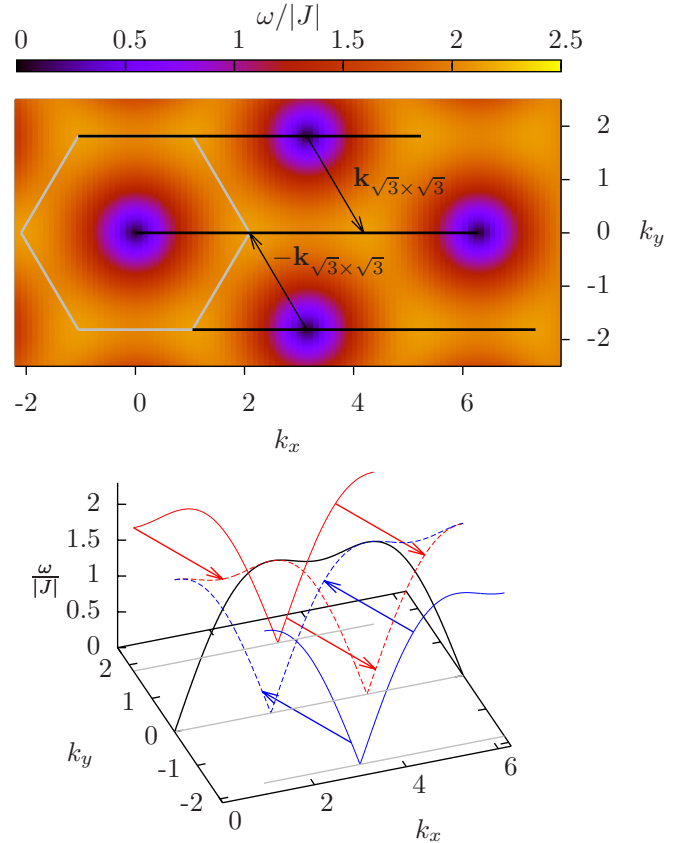}
\caption{\small{\label{fig:freq_shift_10} \emph{Analytical frequencies of the acoustic branch as a function of wave vector $\mathbf k$. In the (10) direction, the shifts by $\mathbf k_{\sqrt3\!\times\!\sqrt3}$ and by $-\mathbf k_{\sqrt3\!\times\!\sqrt3}$ contribute differently, resulting in two apparent branches. The gray hexagon marks the edge of the first Brillouin zone.}}}
\end{center}
\end{figure}

\begin{figure}
\begin{center}
\includegraphics[width=\columnwidth]{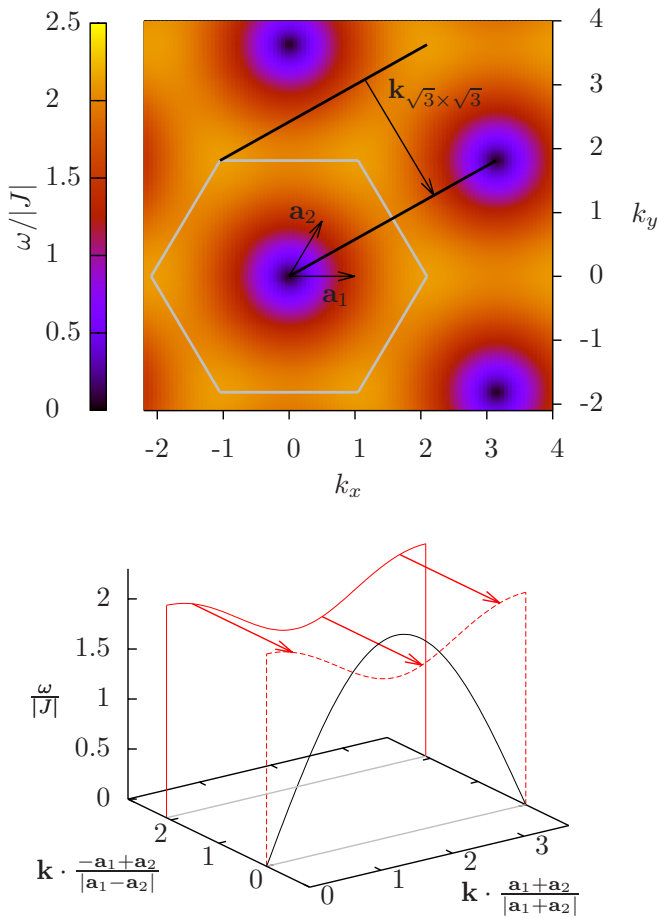}
\caption{\small{\label{fig:freq_shift_11} \emph{Analytical frequencies of the acoustic branch as a function of wave vector $\mathbf k$. In the (11) direction, the shift by $\mathbf k_{\sqrt3\!\times\!\sqrt3}$ creates a fictitious optical branch. The shift by $-\mathbf k_{\sqrt3\!\times\!\sqrt3}$ (not shown) has an identical effect.}}}
\end{center}
\end{figure}

To understand this behavior, we need to know the relationships between the in-plane correlation functions in global and alternative coordinates. In Appendix B we show that
\begin{equation}
C^{xy}_{\mathbf r}(t)\propto \Gamma_{\mathbf r}C^u_{\mathbf r}(t).
\end{equation}
It follows that $C^{xy}_{\mathbf r}(t)$ and $S^{xy}(\mathbf k,\omega)$ depend on the underlying loop structure $\{\sigma\}$ even if $C^u_{\mathbf r}(t)$ does not. For the $\sqrt3\!\times\!\sqrt3$ structure 
the Fourier transform of $\Gamma_{\mathbf r}^{\sqrt3\!\times\!\sqrt3}$ is
\begin{equation}
\hat{\Gamma}^{\sqrt3\!\times\!\sqrt3}({\mathbf k})\propto\sum\limits_{i=1}^{6}\delta({\mathbf k}-{\mathbf k}^i_{\sqrt3\!\times\!\sqrt3}),
\end{equation}
where $\delta$ is Dirac's delta. Using the convolution theorem we find
\begin{equation}
S^{xy}({\mathbf k},\omega)\propto \sum\limits_{i=1}^{6}S^{u}({\mathbf k-\mathbf k^i_{\sqrt3\!\times\!\sqrt3}},\omega).
\end{equation}
This means that modes seen in $S^{u}$ contribute six-fold to $S^{xy}$ as fictitious modes displaced by $ \mathbf k^i_{\sqrt3\!\times\!\sqrt3}$ as shown in Figs.~\ref{fig:freq_shift_10} and \ref{fig:freq_shift_11}. The particular choice of $\mathbf k^i_{\sqrt3\!\times\!\sqrt3}$ is irrelevant unless intensities are considered.Due to the periodicity of $\omega$ each of the three directions will cause the same apparent branches  by shifting the center of the Brillouin zone onto a corner and the opposite corner into the center. In the (10) directions this results in acoustic branches which follow the same dispersion relation as Eq. (\ref{eqn:freq_10}) but shifted by $\pm2\pi/3$. 

In the (11) direction, the apparent optical mode turns out to be a {\emph part of the acoustic branch} and the dispersion relation (for the shifted positions) reads
\begin{eqnarray}
\tilde\omega(l\mathbf e_{11})&=&\omega(l\mathbf e_{11}\pm\mathbf k_{\sqrt3\!\times\!\sqrt3}),\nonumber\\
	&=&\sqrt{\frac72+\cos\left( {\sqrt3}l\right)}.
\label{eqn:freq_11_shift}
\end{eqnarray}

The described behavior is induced by the $\sqrt3\!\times\!\sqrt3$ correlations which in the general case of free $\{\sigma\}$ are present but appear to decay in a power law (Fig.~\ref{fig:sigma_correl}). This means that we have to apply the convolution theorem a second time to describe the shifts in $k$-space if an ensemble of typical loop structures (not pure $\sqrt3\!\times\!\sqrt3$) is considered. Since the Fourier transform of the power law is again a power law, it follows that in principle each mode can be measured at each point in Fourier space with maximal intensity at the position given by the initial shifts $\pm \mathbf k_{\sqrt3\!\times\!\sqrt3}$. This effect causes the washboard pattern by smearing the peaks in k-space; however, as mentioned earlier, the shifted peaks can only be observed for finite systems. We expect no signs of acoustic modes in $S^{xy}({\mathbf k},\omega)$ in the thermodynamic limit.
If, on the other hand, $\sqrt3\!\times\!\sqrt3$ correlations are stabilized by second- and third-nearest neighbor interactions \cite{kagome_berlinsky} the shifted acoustic branch without the washboard would be detectable in experiments and might actually have been observed allready \cite{hyperkagome_exp}.

Considering the described displacements, we find that all observations become consistent. Not only do the positions of maxima in $S^u({\mathbf k},\omega)$ and $S^{xy}({\mathbf k},\omega)$ agree nicely with theory, the deviations caused by non-zero temperature also match. From Fig.~\ref{fig:10_rel} and Fig.~\ref{fig:11_zero} it follows that at the corners of the Brillouin zone, acoustic modes have frequencies above the analytically calculated values while no deviation is noticeable at the midpoints of the zone edges according to Fig.~\ref{fig:10_rel} and Fig.~\ref{fig:11_rel}. These two plots also show that close to the center, frequencies are slightly increased, while they are below theoretical predictions at intermediate distance from the center (for the latter, see also Fig.~\ref{fig:11_zero}).

\section{Conclusions}
For this work we devised a Monte Carlo algorithm that was able to equilibrate the Heisenberg antiferromagnet on the kagome lattice at temperatures more than two orders of magnitude lower than previously reached. To do so, we designed a trial move that allows the ``flipping'' of weather vane loops, and we combined the simulated tempering and heatbath techniques with a modification of the Wang-Landau algorithm. We were able to show that, in agreement with theory, minimal temperature-dependent selection of loop structures occurs over three orders of magnitude in temperature.

Extensive spin dynamics simulations at $k_{\rm B}T_i/|J|=10^{-6}$ provided results that agree very well with analytical calculations: Using suitable coordinates, we observed harmonic modes with the predicted frequencies directly within the spin plane and transverse to it. Considering global Cartesian coordinates, we showed that signals in the dynamic structure factor are shifted due to $\sqrt3\!\times\!\sqrt3$ correlations. Apparent in-plane excitation at the center of the Brillouin zone with non-zero frequency actually belong to acoustic modes at the zone edge. Hence, the classification of these signals as a branch of optical modes by Robert et al. \cite{kagome_sd_prl} was premature.

We argue that due to the decay of the correlations, in-plane excitations will not be measurable for large systems unless next-nearest neighbor interactions stabilize the $\sqrt3\!\times\!\sqrt3$ correlations.


\section{Acknowledgements}
We thank Shan-Ho Tsai for helpful discussions. This work was funded by NSF grant No. DMR-0810223.


\section{Appendix A}
The Wang-Landau algorithm \cite{wanglandau, wanglandau2, wanglandau3} is a method that is designed to adjust a weight function $w_i(Q)$ over time $i$ such that a flat histogram $H(Q)$ can be produced. The parameter $Q$ is often the system energy which allows the determination of thermodynamic quantities such as specific heat and mean energy via the estimation of the density of states as a function of energy. Because we are interested in producing a flat histogram over a logarithmic temperature range, we will treat $Q$ as a general variable in these considerations. In the simplest case, the probability of a configuration $\mathbf X$ is desired to be proportional to $w\left(Q(\mathbf X)\right)$ which means that proposed Monte Carlo moves $\mathbf{X}\rightarrow\mathbf{X}'$ are accepted with probability $P_{\rm acc}(\mathbf{X},\mathbf{X}')=\min\left(1,w(Q(\mathbf X')\right)/w\left(Q(\mathbf X))\right)$. In that way a series of configurations $\mathbf{X}_i$ and values $q_i=Q(\mathbf{X}_i)$ is produced. In order to achieve flatness we modify the weight function after each step:
\begin{equation}
w_{i+1}(Q) = \left\{ \begin{array}{rl}
  \frac{w_{i}(Q)}{f} & \mbox{if $Q=q_{i}$} \vspace{8pt}\\
  w_{i}(Q) & \mbox{otherwise}
\end{array} \right.,
\label{eqn:WL_acc}
\end{equation}
where $f$ is a modification factor which is initially set to $f=e$ (Euler's constant) and gradually reduces to unity, thus causing decreasing changes to $w$. We propose a modification of the original Wang-Landau algorithm which is based on the assumption that especially early in the iteration, the deviations from the detailed balance criterion can be reduced by delaying the modification of the weight function. It is self-evident that a change in $w$ affects the simulation's balance most if it is performed in the proximity of the current position $q_i$ of the system. We, therefore, introduce a delay for the modification of $d$ time steps which allows the simulation to proceed undisturbed, decorrelate, and be less influenced by alterations of $w$. When a modification at $q_{i-d}$ is eventually made, it is necessary to take into account the changes of $w$ at that position between time $i-d$ and $i$. The corrected modification formula reads:
\begin{equation}
w_{i+1}(Q) = \left\{ \begin{array}{rl}
  \frac{w_{i}(Q)}{f^{\frac{w_{i}(q_{i-d})}{w_{i-d}(q_{i-d})}}} & \mbox{if $Q=q_{i-d}$} \vspace{9pt}\\
  w_{i}(Q) & \mbox{otherwise}
\end{array} \right..
\end{equation}
Note that these equations become equal to Eq. \ref{eqn:WL_acc} if $d=0$, i.e., in the case of no delay. To calculate this formula, it is necessary to know $q_{i-d}$ and $w_{i-d}(q_{i-d})$, i.e at each time $i$ the values $q_{i}$ and $w_{i}(q_{i})$ have to be stored for $d$ time steps giving a total of $2d$ numbers to be held in memory. On modern computers, this can be done for large values of $d$; however, we found that a good compromise between increased balance and fast convergence is achieved at $d\approx10^3,10^4$.

It turns out that systematic errors are much smaller in the beginning of the simulations and that the overall convergence is often faster, but in our experience never slower, than in the original Wang-Landau method. In the past this method was used successfully to calibrate one- and two-dimensional weight functions for polymer simulations \cite{my_polymer_cpl,my_polymer_jcp}.

In the present study the weight function $w$ corresponds to the weight factors $W_i$, and the quantity $Q$ is $\log_{10}k_{\rm B}T/|J|$. Modifications of $W$ were performed after each temperature update.

\section{Appendix B}
Consider the product of the $x$-components of two spins as a function of local coordinates ${s}^u$. The three main spin directions lie in the $xy$-plane, while one of which and the $y$-axis include an angle $\phi$ (Fig.~\ref{fig:in_plane_coordinates}).  Instead of taking the average over all possible angles $\phi$, we average only over circular permutations of spin directions $\sigma$, which corresponds to rotations of all spins by angles $\pm\frac23\pi$ around the $z$-axis.

\begin{widetext}
If $\sigma_i=\sigma_j$:
\begin{eqnarray}
\langle s^x_{{\mathbf r}_i}s^x_{{\mathbf r}_j}\rangle&=&\frac13 s^u_{{\mathbf r}_i} s^u_{{\mathbf r}_j}\left( \cos^2 \phi + \cos^2 (\phi+\frac23\pi) + \cos^2 (\phi-\frac23\pi) \right),\nonumber\\
              &=&\frac16 s^u_{{\mathbf r}_i} s^u_{{\mathbf r}_j}\left( 3 + \cos 2\phi + \cos (2\phi+\frac43\pi) + \cos (2\phi-\frac43\pi) \right),\nonumber\\
              &=&\frac12 s^u_{{\mathbf r}_i} s^u_{{\mathbf r}_j}.
\end{eqnarray}
If $\sigma_i\ne\sigma_j$:
\begin{eqnarray}
\langle s^x_{{\mathbf r}_i}s^x_{{\mathbf r}_j}\rangle&=&\frac13 s^u_{{\mathbf r}_i} s^u_{{\mathbf r}_j}\left( \cos\phi \cos(\phi+\frac23\pi) + \cos\phi \cos(\phi-\frac23\pi) + \cos(\phi+\frac23\pi) \cos(\phi-\frac23\pi) \right),\nonumber\\
              &=&\frac16 s^u_{{\mathbf r}_i} s^u_{{\mathbf r}_j}\left( \cos(2\phi+\frac23\pi) + \cos(-\frac23\pi) + \cos(2\phi-\frac23\pi) + \cos(\frac23\pi) + \cos(2\phi) + \cos(\frac43\pi)\right),\nonumber\\
              &=&\frac16 s^u_{{\mathbf r}_i} s^u_{{\mathbf r}_j}\left( \cos(-\frac23\pi) + \cos(\frac23\pi) + \cos(\frac43\pi)\right),\nonumber\\
              &=&-\frac14 s^u_{{\mathbf r}_i} s^u_{{\mathbf r}_j}.
\end{eqnarray}
\end{widetext}

In both cases, the contributions of $\phi$ vanish which means that the average is equivalent to an average over $\phi$.  Furthermore, $\langle s^x_{{\mathbf r}_i}s^x_{{\mathbf r}_j}\rangle_{\{\sigma\}}=\langle s^y_{{\mathbf r}_i}s^y_{{\mathbf r}_j}\rangle_{\{\sigma\}}$ even if $\phi$ is constant because an average over all loop structures $\{\sigma\}$ involves the average over permutations. Following Harris et al.~\cite{kagome_berlinsky} we exploit the fact that collective harmonic excitations $s^u$ are independent of the spin configuration $\{\sigma\}$ and average over all $\{\sigma\}$:
\begin{eqnarray}
\langle s^x_{{\mathbf r}_i}s^x_{{\mathbf r}_j}\rangle_{\{\sigma\}} &=& s^u_{{\mathbf r}_i} s^u_{{\mathbf r}_j}\left(\frac{\tilde \Gamma_{{\mathbf r}_i-{\mathbf r}_j}}2-\frac{1-\tilde \Gamma_{{\mathbf r}_i-{\mathbf r}_j}}4\right),\nonumber\\
                                           &=& s^u_{{\mathbf r}_i} s^u_{{\mathbf r}_j}\left(\frac34 \tilde \Gamma_{{\mathbf r}_i-{\mathbf r}_j}-\frac14\right),
\end{eqnarray}
where $\tilde \Gamma_{{\mathbf r}_i-{\mathbf r}_j}$ denotes the probability for $\sigma_i=\sigma_j$:
\begin{equation}
\tilde \Gamma_{{\mathbf r}_i-{\mathbf r}_j}=\langle \delta_{\sigma_i,\sigma_j}\rangle,
\end{equation}
with $\delta$ being the Kronecker delta. Inserting the aforementioned correlation function $\Gamma$, we obtain
\begin{equation}
\langle s^x_{{\mathbf r}_i}s^x_{{\mathbf r}_j}\rangle_{\{\sigma\}} = \frac34 s^u_{{\mathbf r}_i} s^u_{{\mathbf r}_j} \Gamma_{{\mathbf r}_i-{\mathbf r}_j}.
\end{equation}
It follows that
\begin{equation}
C^{x}_{\mathbf r}(t)=\frac34 \Gamma_{\mathbf r}C^u_{\mathbf r}(t)
\end{equation}
and
\begin{equation}
C^{xy}_{\mathbf r}(t)=C^{x}_{\mathbf r}(t)+C^{y}_{\mathbf r}(t)=\frac32 \Gamma_{\mathbf r}C^u_{\mathbf r}(t).
\end{equation}

\bibliography{refs}{}
\end{document}